\documentstyle[psfig,emulatepasa]{article}

\def\reference{\parskip 0pt\par\noindent\hangindent 0.5 truecm}
\newcommand{\iso}[2]{\hbox{${}^{#1}{\rm #2}$}}

\newcommand{\beqn}{\begin{equation}}
\newcommand{\eeqn}{\end{equation}}
\newcommand{\Msun}{M$_{\odot}$ }
\newcommand{\msun}{M$_{\odot}$}

\newcommand{\mgfour}{$^{24}$Mg }
\newcommand{\mgfive}{$^{25}$Mg }
\newcommand{\mgsix}{$^{26}$Mg }
\newcommand{\mgfiveon}{${^{25}\rm{Mg}}$/${^{24}\rm{Mg}}$  }
\newcommand{\mgsixon}{${^{26}\rm{Mg}}$/${^{24}\rm{Mg}}$  }
\newcommand{\fe}{[Fe/H] }
\begin{document}

\small
\shorttitle{The Chemical Evolution of Magnesium Isotopes}
\shortauthor{Y. Fenner et al.}
%
% Title
% Capitalise the title normally - do not use ALL CAPS.
%

\title{The Chemical Evolution of Magnesium Isotopic Abundances in the
Solar Neighbourhood}

%
% Authors

% **** IMPORTANT: Leave the closing curly bracket line as is. ******

\author{Y. Fenner$^{1,2}$,
 B.~K. Gibson$^{1}$,
 H.-c. Lee$^{1}$,\\
 A.~I. Karakas$^{3}$, 
 J.~C. Lattanzio$^{3}$, \\
 A. Chieffi$^{1,3,4}$,
 M. Limongi$^{1,3,5}$ \&
 D.~Yong$^{6}$ \\
} % IMPORTANT: leave this curly bracket as the first character of this line.

% Date - leave this blank.
\date{}

%%%%%%%%%%% Emulate PASA style %%%%%%%%%%%%%%%%%
%\onecolumn[
\maketitle
\vspace{-20pt}
\small
%%%%%%%%%%% Emulate PASA style %%%%%%%%%%%%%%%%%5

%\maketitle

% Institutions

{\center
$^1$Centre for Astrophysics \& Supercomputing,
Swinburne University, Hawthorn, Victoria, 3122, Australia\\[3mm]
$^2$UCO/Lick Observatory, University of California, Santa Cruz,
Santa Cruz, California, 95064, USA\\[3mm]
$^3$Centre for Stellar \& Planetary Astrophysics, Monash University, Victoria, 
3800, Australia\\[3mm]
$^4$Istituto di Astrofisica Spaziale e Fisica Cosmica, Via Fosso del Cavaliere,
I-00133, Roma, Italy\\[3mm]
$^5$Istituto Nazionale di Astrofisica - Osservatorio Astronomico di Roma, Via
Frascati 33, I-00040, Monteporzio Catone, Italy\\[3mm]
$^6$Department of Astronomy, University of Texas, Austin, TX 78712, USA\\[3mm]
}

% Abstract
%\begin{abstract}

%%%%%%%%%%% Emulate PASA style %%%%%%%%%%%%%%%%%
\begin{center}
{\bfseries Abstract}
\end{center}
\begin{quotation}
\begin{small}
\vspace{-5pt}
%%%%%%%%%%% Emulate PASA style %%%%%%%%%%%%%%%%%

The abundance of the neutron-rich magnesium isotopes observed in
metal-poor stars is explained quantitatively with a chemical evolution
model of the local Galaxy that considers - for the first time - the
metallicity-dependent contribution from intermediate mass stars.
Previous models that simulate the variation of Mg isotopic ratios with
metallicity in the solar neighbourhood have attributed the production
of \iso{25}{Mg} and \iso{26}{Mg} exclusively to hydrostatic burning in
massive stars.
% (e.g. Timmes et al. 1995; Alibes et al. 2001; Goswami \& Prantzos
% 2000).
These models match the data well for \fe $>$ $-$1.0 but severely
underestimate \iso{25,26}{Mg}/\iso{24}{Mg} at lower metallicities.
Earlier studies have noted that this discrepancy may indicate a
significant role played by intermediate-mass stars. Only recently have
detailed calculations of intermediate-mass stellar yields of \mgfive
and \iso{26}{Mg} become available with which to test this hypothesis.
In an extension of previous work, we present a model that successfully
matches the Mg isotopic abundances in nearby Galactic disk stars
through the incorporation of nucleosynthesis predictions of Mg
isotopic production in asymptotic giant branch stars.

%%%%%%%%%%% Emulate PASA style %%%%%%%%%%%%%%%%%
%\end{abstract}
%%%%%%%%%%% End Emulate PASA style %%%%%%%%%%%%%%%%%
%\\


%%%%%%%%%%% Emulate PASA style %%%%%%%%%%%%%%%%%
\end{small}
\end{quotation}
%]
%%%%%%%%%%% End Emulate PASA style %%%%%%%%%%%%%%%%%

\bigskip

\section{Introduction}

Magnesium is one of the few elements for which the stellar abundance
of individual isotopes can be reliably measured. The relative
abundances of Mg isotopes provide a useful probe into the
nucleosynthesis history of the Milky Way because they have their
origin in different classes of stars.  According to standard theories
of stellar evolution, most of the Mg isotopes originate from massive
stars. The heavy Mg isotopes behave as secondary elements inside
massive stars and their production scales with the number of initial
``seed'' nuclei. Consequently, very little \iso{25,26}{Mg} is
synthesised by massive stars until an initial \fe of $\sim$~$-$1 is
reached, whereas the generation of \mgfour operates fairly
independently of initial metallicity.

\mgfive and \iso{26}{Mg} are detected in metal-poor stars in higher
proportions than one would expect if these neutron-rich isotopes
originated exclusively from massive stars (Gay \& Lambert
2000). Detailed models of the evolution of Mg isotopic ratios find
that massive stars alone are insufficient to account for the values of
\iso{25,26}{Mg}/\iso{24}{Mg} at low [Fe/H], hinting at an additional
production site (Timmes et al.~1995; Alibes et al. 2001; Goswami
\& Prantzos 2000).

Karakas \& Lattanzio (2003a,b) have shown that \mgfive and \mgsix
production is substantial in metal-poor intermediate-mass stars
(IMS). In the low-metallicity regime, asymptotic giant branch (AGB)
stars are believed to generate \iso{25}{Mg} and \iso{26}{Mg} from
alpha capture onto $^{22}$Ne triggered by He-shell thermal
pulsing. More massive AGB stars (4~$<$~$m$/\Msun$<$~6) are less common
than lower mass stars but may be a significant production site for
\iso{25}{Mg} and \iso{26}{Mg}.  Temperatures at the base of the
convective envelope in these stars can be high enough to burn
\iso{24}{Mg} via hot bottom burning (HBB) as well as synthesise large
amounts of \iso{25}{Mg} and \iso{26}{Mg}. We explore the possibility
that AGB stars produce sufficient quantities of \mgfive and \mgsix to
resolve the discrepancy between observations and previous model
predictions.

\section{The Chemical Evolution Model}

The temporal and radial evolution of the isotopic abundances in the
Milky Way was calculated under the assumption that the Galaxy formed
via the accretion of gas at a rate decreasing exponentially with time.
For the sake of simplicity, we assumed only a single episode of
primordial gas accretion, with a timescale of 7\,Gyr at the solar
radius. However the results were not significantly different for a
two-phase accretion model. We traced the chemical elements through the
ongoing cycles of star formation, nucleosynthesis, and ejection into
the interstellar medium (ISM) via supernovae (SNe) explosions and
stellar winds. In order to precisely monitor the abundances of
isotopes with different production sites, mass- and
metallicity-dependent stellar lifetimes and yields were employed.

The rate of star formation in this model varies with the square of the
gas surface density and inversely with Galactocentric radius. This
type of radially-dependent law is motivated by the idea that spiral
arm patterns trigger star formation (e.g. Prantzos \& Silk 1998). The
mass distribution of each new generation of stars was governed by the
Kroupa, Tout \& Gilmore (1993) three-component initial mass function
(IMF), with lower and upper mass limits of 0.8 and 100~M$_{\odot}$,
respectively.

Three basic models were constructed, differing only in the adopted
nucleosynthesis prescriptions. Firstly, \emph{LSC + AGB} refers to a
combination of low and intermediate-mass stellar yields from Karakas
\& Lattanzio (2003a,b) and a grid of mass and metallicity dependent (Z
= 0, 10$^{-3}$ \& 0.02) massive star yields provided by Limongi (2001,
unpublished), calculated using the FRANEC code described in Limongi,
Straneiro \& Chieffi (2000) and Limongi \& Chieffi (2002). The second
model, \emph{LSC no AGB}, is identical to the first model but ignores
the AGB contribution to \mgfive and \mgsix production.  Finally, the
model \emph{WW no AGB} replicates \emph{LSC no AGB} but using
metallicity-dependent Woosley \& Weaver (1995) yields for massive
stars. All models adopt yields for Type~Ia SNe from the W7 model of
Iwamoto et al. (1999). The SNe~Ia contribution to chemical evolution
was calculated following the method from Matteucci \& Greggio (1986).
It was assumed that 3\% of binaries culminate in SNe~Ia, since this
fraction provides a good fit to the present-day SNe~Ia rate (e.g.
Alibes et al. 2001).  For stars whose metallicity lies below (above)
the range covered by the nucleosynthesis models we estimate their
yields by extrapolating from the two lowest (highest) metallicity
grids.

The Karakas \& Lattanzio (2003a,b) stellar models comprise a grid of
Mg isotopic yields covering a range of low to intermediate stellar
mass (1~$\le$~$m$/M$_{\odot}$~$\le$~6) and a variety of compositions (Z
= 0.004, 0.008 \& 0.02, supplemented by an unpublished 0.0001 grid
calculated with the same code) that is well-suited for chemical
evolution models. These models have been evolved from the pre-main
sequence to near the end of the thermal-pulsing AGB phase. The
nucleosynthesis calculations are performed separately to determine the
production of the isotopes.

\section{Results \& Discussion}  \label{section:results}

Figure 1 shows the predicted evolution of magnesium isotopic ratios
with [Fe/H] from the models \emph{LSC + AGB} (\emph{solid line}) and
\emph{LSC no AGB} (\emph{dotted line}) for the solar region.
\mgfiveon and \mgsixon are shown in the upper and lower panels,
respectively, along with measured abundance ratios in local dwarfs
from Gay \& Lambert (2000) and cool subdwarfs from Yong (2003).
Representative observational errors are indicated by the large
crosses. Although the quoted errors for \mgfive and \mgsix are
identical in both studies, \mgsix is expected to be more accurately
determined than \mgfive because the $^{26}$MgH line is less blended
with $^{24}$MgH. Both models shown in Figure 1 predict ratios larger
than solar (indicated by squares) but consistent with the data of Gay
\& Lambert. It is not surprising that the models reach similar
present-day values irrespective of whether AGBs are included, because
massive stars are responsible for most of the neutron-rich Mg isotopes
in the present-day ISM. In this model 10\% of \mgsix present in the
ISM at [Fe/H]~=~0 comes from AGB stars compared with $\sim$~70\% at
[Fe/H]~=~$-$1 and nearly 90\% at [Fe/H]~=~$-$2.

\begin{figure}[!p] 
\begin{center}
\psfig{file=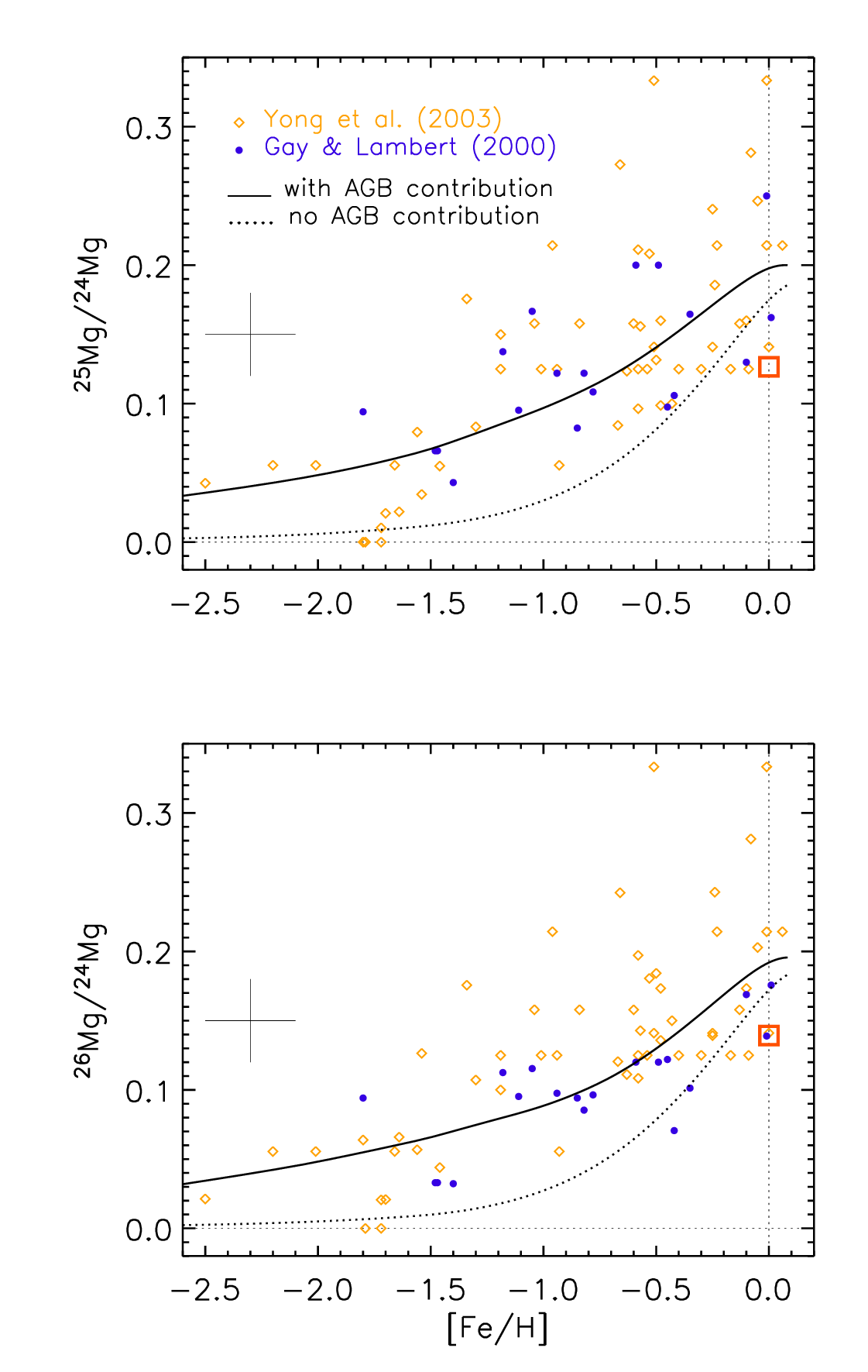,height=17.0cm,angle=0}

\caption{Variation of magnesium isotopic ratios with
    metallicity [Fe/H]. \mgfiveon and \mgsixon are shown in the upper
    and lower panels, respectively. Circles correspond to stellar
    abundances observed by Gay \& Lambert (2000), while diamonds
    represent a sample of halo and thick disk stars from Yong (2003).
    The large cross appearing in both panels indicates typical
    observational errors.  Solar values appear as squares. The
    predicted trend of our solar neighbourhood model incorporating Mg
    isotopic yields from AGBs (\emph{solid line}) is shown against a
    model without the AGB contribution (\emph{dotted line}). Both
    models arrive at similar present-day values, however only the AGB
    model matches the empirical data at low metallicities.}

\label{fig:fig1}
\end{center} 
\end{figure}

The dotted lines reveal that below [Fe/H]~$\sim$~$-$1, massive stars
alone seriously under produce \mgfive and \mgsix with respect to
\iso{24}{Mg}. Much better agreement is obtained by including the
contribution from AGB stars. In particular, most of the heavy Mg
isotopic abundance at low metallicity is controlled by the
4-6~M$_{\odot}$ stars that undergo hot bottom burning and whose
He-shells are hot enough to trigger the
\iso{22}Ne($\alpha,n$)\iso{25}{Mg} and
\iso{22}Ne($\alpha,\gamma$)\iso{26}{Mg} reactions (Karakas \&
Lattanzio 2003a,b). These stars typically have lifetimes between 60
and 170~Myr.

\begin{figure}[!t] 
\begin{center}
\psfig{file=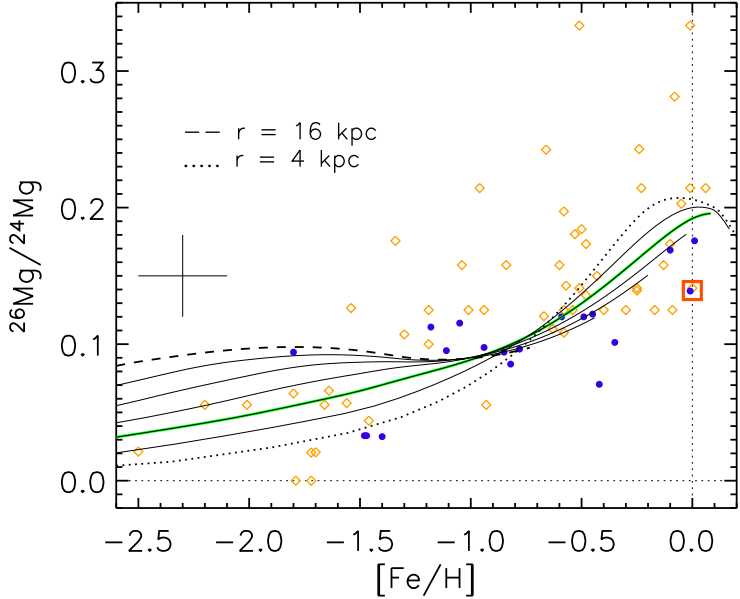,height=9.0cm,angle=0}

\caption{Evolution of \mgsixon as a function of \fe at different
Galactic radii. Symbols have the same meaning as in Fig 1. The curves
correspond to predicted behaviour at seven evenly spaced radii,
ranging from 4-16~kpc for the \emph{LSC + AGB} model. The green line
indicates the solar radius (8~kpc). }

\label{fig:fig2}
\end{center} 
\end{figure}
\begin{figure}[!t] 
\begin{center}
\psfig{file=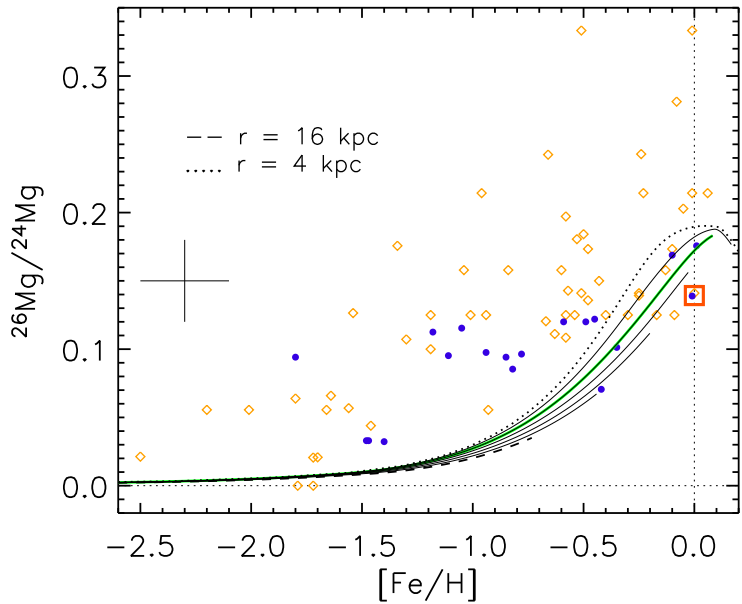,height=9.0cm,angle=0}

\caption{Same as Fig 2. but for model \emph{LSC no AGB} (see text for
details). }

\label{fig:fig3}
\end{center} 
\end{figure}

The high abundances of \mgfive and \mgsix from the Yong (2003)
data-set pose more of a challenge and are difficult to reconcile with
our models, even with an AGB contribution. These stars were selected
by their large transverse velocities as part of an effort to compile a
sample of metal-poor unevolved subdwarfs (Yong \& Lambert 2003). Based
on their high reduced proper motion, it is likely that most of the
sample is distinct from the thin disk, belonging instead to the halo
at low metallicities or the thick disk at higher metallicities. One
might expect that stars enhanced in \mgfive and \mgsix are also either
1) rich in s-process elements if they have been heavily contaminated
by AGB stars; or 2) $\alpha$-enhanced if they belong to the halo/thick
disk. Yong (2003) reports that there does not appear to be an obvious
relationship between high values of \iso{25,26}{Mg} and the abundance
of s-process or $\alpha$-elements.

The predicted trend of \mgsixon with \fe is plotted in Figure 2 at
different Galactic radii for the \emph{LSC + AGB} model. The dotted
curve corresponds to the innermost radius of the model (4~kpc) and the
dashed curve to the outermost radius (16~kpc). The exponential
timescale for infalling gas was assumed to increase linearly with
radius, taking on a value of 7~Gyr at the solar radius. Due to this
``inside-out'' accretion scenario and the radially-dependent star
formation law, the timescale of SF is much longer in the outer disk.
Only \mgsixon is plotted because the \iso{25}{Mg}H line is less
reliably determined than the \iso{26}{Mg}H line due to more severe
blending with the \iso{24}{Mg}H feature (Gay \& Lambert 2000). This
model predicts a greater spread in \mgsixon in metal-poor stars versus
metal-rich stars throughout the Galaxy. The dispersion in the
metal-poor regime reflects the dependence of \mgfive and \mgsix
abundance on the timescale of star formation. Since the lowest
metallicity AGB models predict the highest \mgfive and \mgsix yields,
environments such as the outer disk, with protracted star formation
and slowly increasing metallicity over time, encourage generations of
IMS to elevate \mgfiveon and \iso{26}{Mg}/\iso{24}{Mg}. The inner
Galaxy by contrast, is believed to have experienced intense star
formation very early in its history, rapidly enriching the ISM with
metals. Once [Fe/H] exceeds about $-$1, massive stars begin to surpass
IMS as the chief source of neutron-rich Mg isotopes. The spread
becomes smaller for [Fe/H]~$>$~$-$1, at which point metallicity rather
than star formation history drives \iso{26}{Mg}/\iso{24}{Mg}.

Figure 3 illustrates the lack of dispersion across different radii for
the \emph{LSC no AGB} model. There is some scatter in \mgsixon at high
[Fe/H] that stems from the delayed release of iron from Type~Ia SNe.
As most of the ejecta from SNe~Ia is in the form of iron, these events
increase [Fe/H] to a greater extent than overall metallicity. A
consequence of the characteristic SNe~Ia time delay of about 1~Gyr is
that \emph{for the same abundance of iron}, the inner disk is expected
to be $\alpha$-enhanced and therefore have greater total abundance of
nuclei than the outer disk. With more seed nuclei, the inner disk for
a given value of [Fe/H] should see higher values of
\iso{26}{Mg}/\iso{24}{Mg} than the outer disk. Precisely this effect
is revealed in Figure 3.

\begin{figure}[!tb] 
\begin{center}
\psfig{file=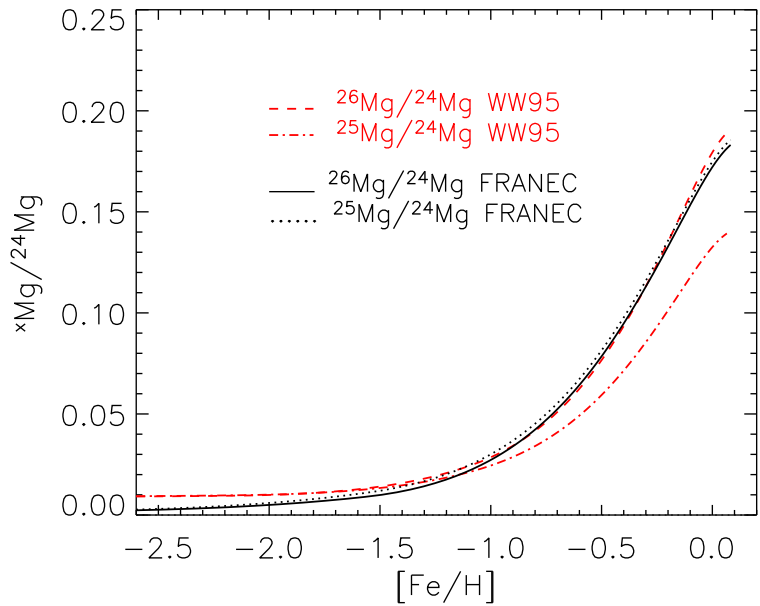,height=9.0cm,angle=0}

\caption{Evolution of magnesium isotopic ratios for models using
different nucleosynthetic prescriptions for massive star yields. Red
lines denote massive star yields from Woosley \& Weaver (1995) and
black lines denote yields from Limongi et~al. (2000; 2002). The AGB
contribution has been ignored for the sake of comparison. There is
good agreement between the two sets of yields.}

\label{fig:fig4}
\end{center} 
\end{figure}

For the sake of comparing how robust our results are to explosive SNe
yields from different authors, we plot in Figure~4 the predicted
evolution of \mgfiveon and \mgsixon for the solar neighbourhood for
the \emph{LSC no AGB} and \emph{WW no AGB} models. There is excellent
agreement between the two models, although Woosley \& Weaver's (1995)
nucleosynthesis models predict about 20\% less \mgfive than \mgsix,
while Limongi et~al. (2000,2002) produce these isotopes in roughly
equal numbers. The Mg isotopic yields depend on the adopted
\iso{12}{C}($\alpha$,$\gamma$)\iso{16}{O} rate (Imbriani et~al. 2001).
Specifically, lowering the value of
\iso{12}{C}($\alpha$,$\gamma$)\iso{16}{O} leads to higher final yields
of \iso{24}{Mg} but leaves the heavier Mg isotopes relatively
uneffected. The Limongi et~al. yields employed in this Milky Way model
use a large \iso{12}{C}($\alpha$,$\gamma$)\iso{16}{O} rate, hence
producing low \iso{24}{Mg} yields. The \mgfiveon and \mgsixon ratios
could be decreased with a lower
\iso{12}{C}($\alpha$,$\gamma$)\iso{16}{O} value such as that from Kunz
et~al. (2002) which is used in the latest version of the FRANEC code
(Limongi \& Chieffi 2003). In this case, the Mg isotopic ratios could
be reconciled with solar values, but at the expense of obtaining a
good fit to the Gay \& Lambert (2000) data.

\begin{figure}[!htb] 
\begin{center}
\psfig{file=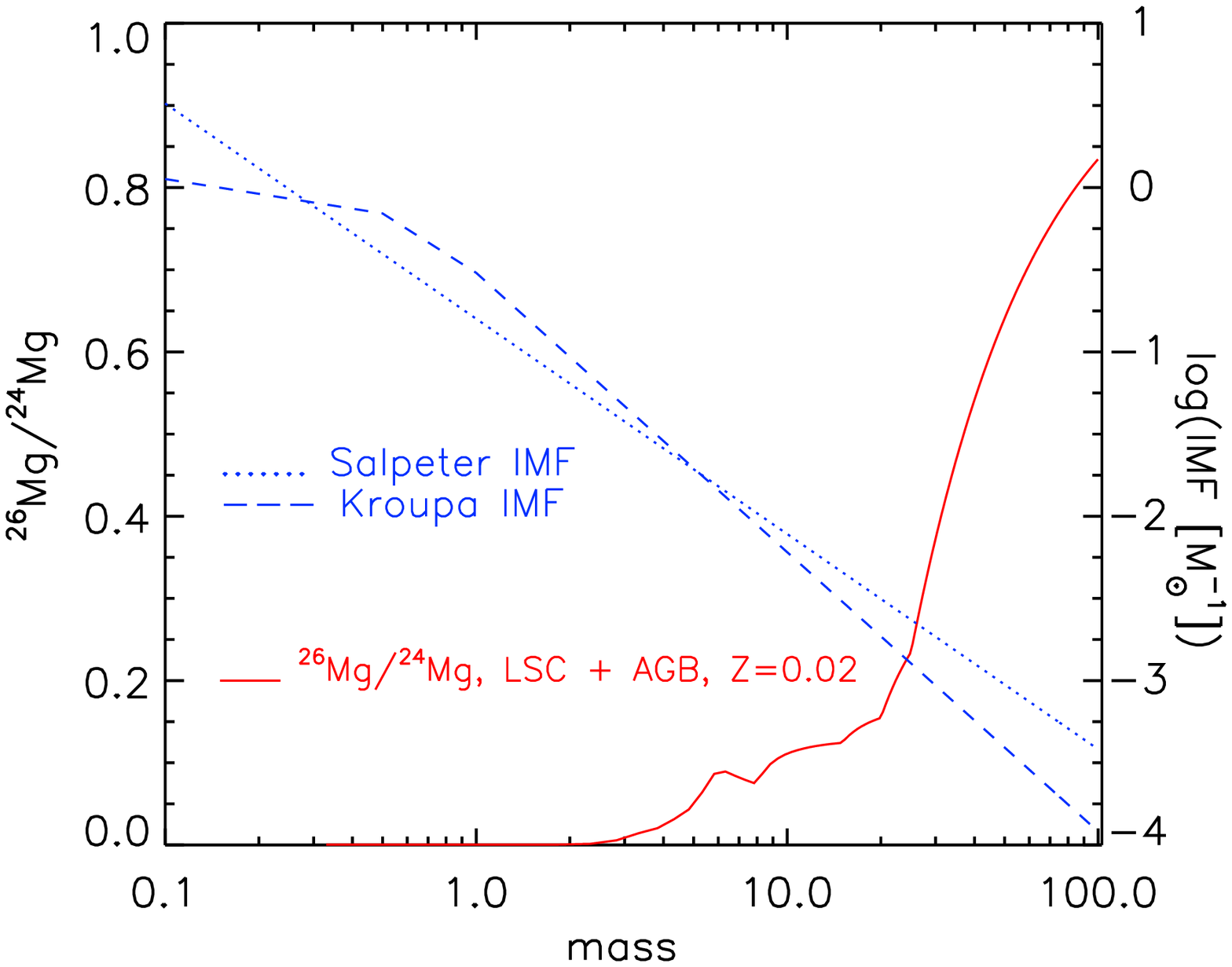,height=9.0cm,angle=0}

\caption{Yield of \mgsixon vs initial stellar mass for solar
metallicity nucleosynthesis models (\emph{solid line}). Yields for
low and intermediate-mass stars are from Karakas \& Lattanzio (2003a,b)
and from Limongi et al. (2000; 2002) for massive stars. The dotted
line corresponds to the Salpeter (1955) initial mass function (IMF) and
the dashed line indicates the Kroupa, Tout \& Gilmore (1993) IMF. Both
functions are normalised to unity over the mass range 0.08-100~\Msun
(the scale appears on the right-hand vertical axis).}

\label{fig:fig5}
\end{center} 
\end{figure}

One of the greatest uncertainties in the model concerns the precise
form of the stellar IMF. While most elemental yields are fairly robust
to changes in the slope and limits of the IMF, this is not the case
for the heavy Mg isotopes, whose production in solar metallicity
massive stars rises very sharply with stellar mass. Thus, the most
massive stars play a crucial role in determining \mgfiveon and
\iso{26}{Mg}/\iso{24}{Mg} despite being vastly outnumbered by lower
mass stars. Figure~5 displays the yield of \mgsix relative to \mgfour
as a function of initial mass for the solar metallicity models of
Karakas \& Lattanzio (2003a,b) and Limongi et al. (2000,2002)
(\emph{solid line}). The solar metallicity massive star yields extend
only as far as 35 \Msun and were linearly extrapolated up to
100~M$_{\odot}$. The largest mass in the AGB models is 6~\msun. For
most elements, including Fe and \iso{24}{Mg}, the yields between
6-8~\Msun were derived by extrapolating up from the AGB models.
However we conservatively assumed that the 6-8~\Msun stars eject the
same quantity of \mgfive and \mgsix as the 6~\Msun star. This causes a
dip in \mgsixon at 8~\msun. Superimposed on the same figure are IMFs
from Salpeter (1955) (\emph{dotted line}) and Kroupa, Tout \& Gilmore
(1993) (\emph{dashed line}) normalized over the mass range
0.08-100~M$_{\odot}$. The Salpeter IMF has the form of a single slope
power-law (this model uses an index of 1.31) and places a higher
proportion of the mass of a stellar generation into both the lower and
upper extremes of the mass distribution when compared with the Kroupa
et~al. (1993) IMF. It is apparent that the role of stars between
0.3~$\ge$~$m$/M$_{\odot}$~$\ge$~6 is emphasised by adopting a Kroupa
et~al.  IMF, whereas the frequency of high mass stars is increased
with the Salpeter IMF.

One can anticipate that a Kroupa et~al. (1993) IMF would give rise to
more AGB stars than a Salpeter (1955) function, leading to higher
values of \mgsixon at low metallicity. At metallicities approaching
solar however, the Salpeter IMF should generate the highest \mgsixon
ratios, since it favours the birth of massive stars when compared with
the Kroupa et~al. law. The influence of the IMF can be seen by
comparing the results presented in Figure~6, which were derived using
the Salpeter function, with those from Figure~2. A ratio of
\mgsixon$\sim$0.3 at [Fe/H]~=~0 is obtained with a Salpeter IMF. This
is $\sim$~50\% higher than in the Kroupa et~al.  case and over twice
the solar value. Owing to the production of fewer AGB stars, \mgsixon
at [Fe/H]~=~$-$2 in the Salpeter case is about half the value derived
with the Kroupa IMF. The Kroupa et~al. model is in better agreement
with the Gay \& Lambert (2000) data set. Although the large values of
\mgsixon measured by Yong (2003) in higher metallicity stars could be
attained by increasing the role of massive stars, this comes at the
expense of satisfying observations at low [Fe/H]. Multi-component IMFs
with steeper slopes at high mass are favoured over the Salpeter single
power law on both observational and theoretical grounds (e.g. Kroupa
et~al. 1993; Scalo 1986).

\begin{figure}[!htb] 
\begin{center}
\psfig{file=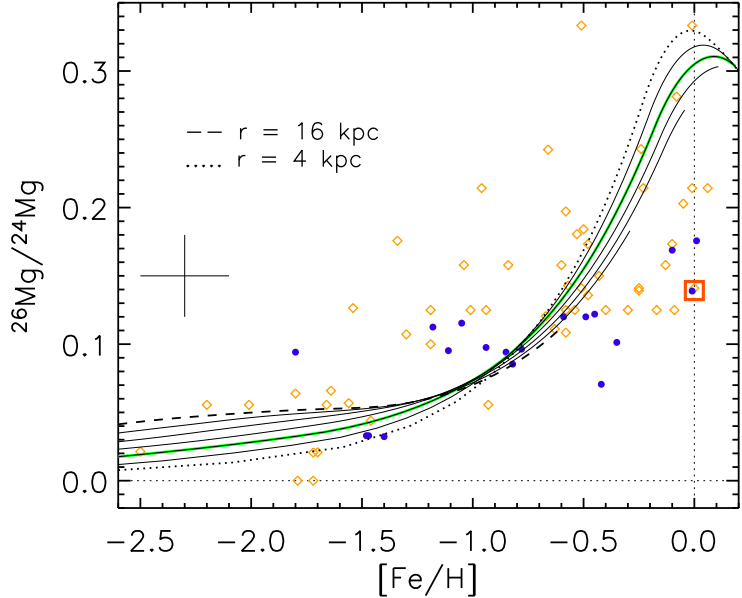,height=9.0cm,angle=0}

\caption{Same as Fig 2. but for the case of a single component
Salpeter (1955) initial mass function in place of Kroupa, Tout \&
Gilmore (1993).}

\label{fig:fig6}
\end{center} 
\end{figure}

It has been suggested that the mass distribution of a stellar
generation is influenced by factors including the thermal energy and
chemical composition of the star-forming gas (Larson
1998). Accordingly, the IMF might be expected to evolve over
time. There are theoretical arguments that the IMF of primordial gas
would be biased toward higher mass stars, while higher metallicity
environments would form relatively more low-mass stars (Kroupa
2001). If the IMF followed this trend then the fit between our model
predictions and empirical constraints would worsen. At present
however, the observational evidence for a variable IMF is not
overwhelming.

\section{Conclusions}

Thermally-pulsing AGB stars are shown to be excellent candidates for
the additional production site of neutron-rich Mg isotopes needed to
account for observations at low metallicities. The failure of previous
chemical evolution models to match \mgfiveon and \mgsixon observed in
local metal-poor stars may be resolved if intermediate-mass stars
produce \mgfive and \mgsix in quantities given by the calculations of
Karakas \& Lattanzio (2003a,b).

According to the chemical evolution model presented in this paper,
massive stars are responsible for most of the heavy Mg isotopes in the
present-day ISM, but played a secondary role to 4-6~\Msun AGB stars at
earlier epochs. A consequence of this model is that the spread in
\mgfiveon and \mgsixon should be greater at low metallicities across
the Milky Way, although this conclusion was shown to be sensitive to
the adopted initial mass function. While the high ratios of \mgfiveon
and \mgsixon in Yong's (2003) halo and thick disk sample remain a
mystery and warrant further investigation, our model provides an
excellent match to the measurements of local stars from Gay \& Lambert
(2000).

\section*{Acknowledgments}
The financial support of the Australian Research Council
(through its Large Grant, Discovery Project, and Linkage International
schemes) and the Victorian Partnership for Advanced Computing (through its
Expertise Grants scheme) is acknowledged.

\section*{References}

\reference Alibes, A., Labay, J. \& Canal, R., 2001, A\&A, 370, 1103
\reference Gay, P. L. \& Lambert, D. L., 2000, ApJ, 533, 260
\reference Goswami, A. \& Prantzos, N., 2000, A\&A, 359, 191
\reference Imbriani, G., Limongi, M., Gialanella, L., Terrasi, F.,
        Straniero, O. \& Chieffi, A., 2001, ApJ, 558, 903
\reference Iwamoto, K., Brachwitz, F., Nomoto, K., Kishimoto, N.,
        Umeda, H., Hix, W. R. \& Thielemann, F.-K., 1999, ApJS, 125, 439
\reference Karakas, A. I. \& Lattanzio, J. C., 2003a, Carnegie Observatories 
Astrophysics Series, Vol. 4: Origin and Evolution of the Elements, ed. 
A. McWilliam and M. Rauch (Pasadena: Carnegie Observatories,\hfill\break
\tt http://www.ociw.edu/ociw/symposia/series/symposium4/proceedings.html\rm) 
\reference Karakas, A. I. \& Lattanzio, J. C., 2003b, PASA, in press
\reference Kroupa, P., Tout, C. A. \& Gilmore, G., 1993, MNRAS, 262, 545
\reference Kroupa, P., 2001, MNRAS, 322, 231
\reference Kunz, R., Fey, M., Jaeger, M., Mayer, A., Hammer, J. W.,
        Staudt, G., Harissopulos,~S. \& Paradellis, T., 2002, ApJ, 567, 643
\reference Larson, R. B., 1998, MNRAS, 301, 569
\reference Limongi, M., Straniero, O. \& Chieffi, A., 2000, ApJS, 129, 625
\reference Limongi, M. \& Chieffi, A., 2002, PASA, 19, 246
\reference Limongi, M. \& Chieffi, A., 2003, ApJ, in press (astro-ph/0304185)
\reference Matteucci, F. \& Greggio, L., 1986, A\&A, 154, 279 
\reference Prantzos, N. \& Silk, J., 1998, ApJ, 507, 229
\reference Salpeter, E. E., 1955, ApJ, 121, 161
\reference Scalo, J. M., 1986, Fund. Cosm. Phys., 11, 1 
\reference Timmes, F. X., Woosley, S. E. \& Weaver, T. A., 1995, ApJS, 98, 617
\reference Woosley, S. E. \& Weaver, T. A. 1995, ApJS, 101, 181
\reference Yong, D., 2003, in preparation
\reference Yong, D. \& Lambert, D. L., 2003, PASP, 115, 22

\end{document}